\documentclass[prd,nofootinbib,letterpaper,floatfix,superscriptaddress,preprintnumbers]{revtex4}
\usepackage{color}

\usepackage{graphicx}

\usepackage{amsmath}
\usepackage{amssymb}
\usepackage{subfigure}
\usepackage{url}

\usepackage{multirow}
\usepackage{threeparttable}
\usepackage{paralist}

\usepackage{color}
\definecolor{darkgreen}{rgb}{0,0.5,0}

\newcommand{\eff}{\text{eff}}

\DeclareRobustCommand{\Sec}[1]{Sec.~\ref{#1}}

\DeclareRobustCommand{\Fig}[1]{Fig.~\ref{#1}}

\DeclareRobustCommand{\Eq}[1]{Eq.~(\ref{#1})}

\DeclareRobustCommand{\Ref}[1]{Ref.~\cite{#1}}
\DeclareRobustCommand{\Refs}[1]{Refs.~\cite{#1}}

\newcommand{\be}{\begin{equation}}
\newcommand{\ee}{\end{equation}}

\newcommand{\mb}[1]{\boldsymbol{#1}}

\newcommand{\kahler}{K\"{a}hler }

\def\Tr{\mathop{\rm Tr}}

\begin{document}

\title{The New Flavor of Higgsed Gauge Mediation}

\author{Nathaniel Craig}
\email{ncraig@ias.edu}
\affiliation{Institute for Advanced Study, Princeton, NJ 08540, USA \\
Department of Physics and Astronomy, Rutgers University, Piscataway, NJ 08854, USA}

\author{Matthew McCullough}
\email{mccull@mit.edu}
\affiliation{Center for Theoretical Physics, Massachusetts Institute of Technology, Cambridge, MA 02139, USA}

\author{Jesse Thaler}
\email{jthaler@mit.edu}
\affiliation{Center for Theoretical Physics, Massachusetts Institute of Technology, Cambridge, MA 02139, USA}

\date{\today}

\begin{abstract}
Recent LHC bounds on squark masses combined with naturalness and flavor considerations motivate non-trivial sfermion mass spectra in the supersymmetric Standard Model.  These can arise if supersymmetry breaking is communicated to the visible sector via new extended gauge symmetries.  Such extended symmetries must be spontaneously broken, or confined, complicating the calculation of soft masses.  We develop a new formalism for calculating perturbative gauge-mediated two-loop soft masses for gauge groups with arbitrary patterns of spontaneous symmetry breaking, simplifying the framework of ``Higgsed gauge mediation.''  The resulting expressions can be applied to Abelian and non-Abelian gauge groups, opening new avenues for supersymmetric model building. We present a number of examples using our method, ranging from grand unified threshold corrections in standard gauge mediation to soft masses in gauge extensions of the Higgs sector that can raise the Higgs mass through non-decoupling $D$-terms.  We also outline a new mediation mechanism called ``flavor mediation'', where supersymmetry breaking is communicated via a gauged subgroup of Standard Model flavor symmetries.  Flavor mediation can automatically generate suppressed masses for third-generation squarks and implies a nearly exact $U(2)$ symmetry in the first two generations, yielding a ``natural SUSY'' spectrum without imposing {\it ad hoc} global symmetries or giving preferential treatment to particular generations.

\end{abstract}

\preprint{MIT-CTP {4338}, RU-NHETC {2011-26}}

\maketitle

\section{Introduction}
\label{sec:introduction}
As the LHC explores the high energy frontier, weak-scale supersymmetry (SUSY) remains a compelling solution to the hierarchy problem and a well-motivated target for LHC searches.  The first few inverse femtobarns of LHC data already place strong bounds on the spectrum of superpartners, and these bounds have refocused experimental and theoretical attention towards SUSY scenarios with alternative spectra.

One of the most elegant mechanisms to generate soft masses in the supersymmetric Standard Model (SSM) is gauge mediation (see e.g.\ \cite{Giudice:1998bp} and references therein), where SUSY-breaking messengers with Standard Model (SM) charges induce soft masses through gauge interactions.  A key advantage to gauge mediation is that SSM soft masses are flavor universal, allowing light sfermions to be consistent with stringent flavor bounds.  On the other hand, flavor-universality limits the possible spectra achievable in even the most general forms of gauge mediation \cite{Meade:2008wd}.  In particular, ordinary gauge mediation can never realize ``natural SUSY'' models with light third-generation squarks \cite{Dine:1993np,Pouliot:1993zm,Barbieri:1995rs,Dimopoulos:1995mi,Pomarol:1995xc,Barbieri:1995uv,Cohen:1996vb,Barbieri:1997tu,Gabella:2007cp,Sundrum:2009gv,Barbieri:2010pd,Barbieri:2010ar,Craig:2011yk,Gherghetta:2011wc,Kats:2011qh,Papucci:2011wy,Brust:2011tb,Delgado:2011kr,Desai:2011th,Akula:2011jx,Ajaib:2011hs,Ishiwata:2011fu,Lodone:2011pv,He:2011tp,Arvanitaki:2011ck,Auzzi:2011eu,Csaki:2012fh}, a possibility which has gained recent interest after the 1.1 fb$^{-1}$ dataset.

In this paper, we will highlight the potential importance of extended gauge symmetries for SUSY.  Any new gauge symmetries beyond the SM must be spontaneously broken or confined.  The former naturally leads to models of ``Higgsed gauge mediation'' \cite{Gorbatov:2008qa}, where soft masses depend on two different thresholds:  the masses of messengers and the masses of gauge bosons.  To date, there have been relatively few studies of Higgsed gauge mediation \cite{Cheng:1998nb,Cheng:1998hc,Kaplan:1998jk,Kaplan:1999iq,Everett:2000hb,Dermisek:2006qj,Langacker:2007ac,Gorbatov:2008qa,Langacker:2008yv,Buican:2009vv,Intriligator:2010be,Craig:2011ev},\footnote{Most studies to date do not include the complete threshold corrections due to vector masses, consider only Abelian gauge symmetries, or treat only one-loop contributions that arise when Higgsing fields couple directly to SUSY-breaking fields.  In models of deconstructed gaugino mediation, part of the total soft mass arises due to Higgsed gauge mediation, and expressions which include these contributions, alongside contributions from the unbroken SM gauge group, can be found in \Refs{Auzzi:2010mb,Auzzi:2010xc,Auzzi:2011gh,Auzzi:2011wt,Auzzi:2011eu}.} likely owing to the complicated functional form of two-loop sfermion masses.  In this work, we present a new and simpler formalism for Higgsed gauge mediation, which relies only on a spurion analysis of the effective K\"ahler potential.  

Armed with a more transparent understanding of Higgsed gauge mediation, we then study some interesting applications, including threshold corrections in grand unified theories (GUTs), alternative spectra in deconstructed orbifold GUTs, and soft mass contributions arising when SUSY breaking is communicated by extended gauge symmetries in the SSM Higgs sector.  Such extended gauge symmetries can raise the Higgs mass above the LEP bound (by an amount consistent with current possible hints \cite{ATLAS-CONF-2011-163, CMS-PAS-HIG-11-032}) through the presence of non-decoupling $D$-terms \cite{Batra:2003nj,Maloney:2004rc}.  

Our key example is ``flavor mediation'', which is motivated by natural SUSY models \cite{Dine:1993np,Pouliot:1993zm,Barbieri:1995rs,Dimopoulos:1995mi,Pomarol:1995xc,Barbieri:1995uv,Cohen:1996vb,Barbieri:1997tu,Gabella:2007cp,Sundrum:2009gv,Barbieri:2010pd,Barbieri:2010ar,Craig:2011yk,Gherghetta:2011wc,Kats:2011qh,Papucci:2011wy,Brust:2011tb,Delgado:2011kr,Desai:2011th,Akula:2011jx,Ajaib:2011hs,Ishiwata:2011fu,Lodone:2011pv,He:2011tp,Arvanitaki:2011ck,Auzzi:2011eu,Csaki:2012fh} with light stops and sbottoms.  Models of flavor mediation involve gauged flavor symmetries, and they exhibit the intriguing feature that hierarchical SM fermion masses can lead to a nearly exact $U(2)$ flavor symmetry in the squarks.  Unlike other approaches to ``flavorful SUSY'' (e.g. \cite{Dine:1993np,Pouliot:1993zm,Pomarol:1995xc,Barbieri:1995uv,Barbieri:1997tu,Gabella:2007cp,Sundrum:2009gv,Craig:2011yk,Gherghetta:2011wc,Brust:2011tb,Delgado:2011kr,Auzzi:2011eu,Csaki:2012fh}), this $U(2)$ flavor symmetry arises without having to impose additional symmetries by hand and without having to treat the first two generations preferentially over the third.  Here, we sketch some basic features of flavor mediation, leaving a full study to future work.

The remainder of this paper is organized as follows.  In the next section, we show how the spectrum of Higgsed gauge mediation can be derived using the two-loop effective K\"ahler potential.  We present three applications of our method in \Sec{sec:examples}, and sketch the structure of flavor mediation in \Sec{sec:exampleFlav}.  We conclude in \Sec{sec:conclude}.

\section{Higgsed Gauge Mediation}
\label{sec:HGM}

Higgsed gauge mediation arises when messenger fields, charged under a spontaneously-broken gauge group, have superpotential couplings to SUSY-breaking fields.  As in unbroken gauge mediation, the resulting non-holomorphic sfermion soft masses can be explicitly calculated in components by performing a two-loop Feynman diagram computation as in \Ref{Gorbatov:2008qa}. This captures the full form of the two-loop soft masses, albeit at the cost of simplicity.  Here, we give a more transparent derivation of the sfermion masses by studying corrections to the \kahler potential arising after integrating out the gauge and messenger superfields. This approach yields the sfermion masses to leading order in the ratio of SUSY-breaking and messenger scales $F/M$, and to all orders in the ratio of vector and messenger scales $M_V^2/M^2$.  Corrections at higher order in $F/M^2$ are not included in this calculation, however these corrections are small for $F/M^2 \lesssim 0.8$, as shown in \Ref{Gorbatov:2008qa}.  Since vacuum stability requires $F/M^2 < 1$, the leading-order expressions obtained herein are valid in the majority of parameter space. 
 In addition, as we will find, the result at leading order in $F/M^2$ is relatively compact and straightforward to implement, whereas the full result including higher-order corrections in \Ref{Gorbatov:2008qa} is somewhat more cumbersome.

For unbroken gauge mediation, sfermion soft masses can be found by analytically continuing the squark wavefunction renormalization \cite{Giudice:1997ni,ArkaniHamed:1998kj}.  For Higgsed gauge mediation, though, na\"{i}ve analytic continuation does not capture the full effect of Higgsing because the gauge boson masses constitute a supersymmetric threshold. Thus we will instead consider the full effective \kahler potential, which by definition includes appropriate threshold matching.  The one-loop effective \kahler potential \cite{Grisaru:1996ve} is a useful tool for studying the generation of soft masses for scalars when such masses arise at one-loop order.  In gauge mediation, sfermion masses are first generated at two loops, forcing us to consider the {\it two-loop} effective \kahler potential.\footnote{Throughout this paper, we assume that the leading sfermion masses arise at two loops, requiring that one-loop $D$-term contributions vanish as a result of vanishing traces or messenger parity.  Also, whenever the messengers also break the gauge symmetry, the dominant sfermion masses arise at one loop, as has been considered in e.g.\ \Ref{Buican:2009vv} or in more detail in \Ref{Intriligator:2010be}.  However, this is specific to the case where the Higgsing fields couple directly to SUSY breaking.  This is rather restrictive and does not cover scenarios where it may be desirable to charge the messengers under multiple gauge groups, both broken and unbroken.  In addition, these one-loop contributions can be tachyonic, which may be problematic, so we do not consider this scenario here.}  Fortunately, the two-loop effective \kahler potential has been determined for general $\mathcal{N}=1$ SUSY theories in \Ref{Nibbelink:2005wc}, and we employ those results here.

\subsection{Effective \kahler Potential}
\label{sec:Kahler}

Our starting point is a SUSY theory with a $U(1)'$ gauge symmetry with gauge coupling $g'$; we will generalize to non-Abelian groups in \Sec{sec:nonab}.  This gauge group is spontaneously broken in a supersymmetric manner, generating a mass $M_V$ for the vector superfield.   We include messenger chiral superfields $\mb{\Phi}/\mb{\Phi^c}$ which have equal and opposite charges $\pm q_\Phi$ under the $U(1)'$ gauge symmetry.  These messengers couple to a SUSY-breaking superfield $\mb{X}$ with superpotential\footnote{We denote superfields in bold font ($\mb{X}$) and their lowest components in plain font ($X$).}
\be
\mb{W} = \mb{X} \mb{\Phi} \mb{\Phi^c} .
\ee
Throughout, we will treat $\mb{X}$ as a background superfield with vacuum expectation value (vev) $\langle \mb{X} \rangle = M + \theta^2 F$.  Finally, we include visible sector chiral superfields $\mb{q}$ with charge $q_q$ under the $U(1)'$ gauge symmetry.

The two-loop effective \kahler potential is a function of the messenger masses $|\mb{M_\Phi}|^2$ and the vector superfield mass $\mb{M_V}^2$.  Both of the these quantities can be expressed as full superfields
\be
|\mb{M_\Phi}|^2 \equiv \mb{X}^\dagger \mb{X}, \qquad \mb{M_V}^2 \equiv M_V^2 + 2 q_q^2 g'^2 \mb{q}^\dagger \mb{q},
\ee
where we have included the visible sector fields $\mb{q}$ as a background spurion in the vector mass.  This technique for accounting for $\mb{X}$ and $\mb{q}$ is reminiscent of analytic continuation into superspace \cite{Giudice:1997ni,ArkaniHamed:1998kj}, and has the same restriction that we only capture the leading effects in $F/M$.

\begin{figure}[t]
\centering
\includegraphics[height=1.5in]{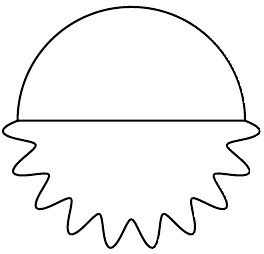}
\caption{The two-loop vacuum diagram contributing to scalar soft masses.  The propagators for the messenger superfields are in the upper half and depend on the messenger mass-squared $|\mb{M_\Phi}|^2=\mb{X}^\dagger \mb{X}$.  The vector superfield propagator on the lower line is a function of the (analytically-continued) mass-squared $\mb{M_V}^2 = M_V^2 + 2 q_q^2 g'^2 \mb{q}^\dagger \mb{q}$ which depends on the Higgsing superfields and the background visible-sector superfields.}
\label{fig:loop}
\end{figure}

To calculate the sfermion soft masses, we simply need to identify terms in the effective \kahler potential that depend on both $\mb{X}^\dagger \mb{X}$ and $\mb{q}^\dagger \mb{q}$.  Examining the two-loop result from \Ref{Nibbelink:2005wc}, there exists only one such term,
\be
K_{2L} = -2 q_\Phi^2 g'^2 I(|\mb{M_\Phi}|^2,|\mb{M_\Phi}|^2,\mb{M_V}^2),
\label{eq:int}
\ee
where
\be
I(|\mb{M_\Phi}|^2,|\mb{M_\Phi}|^2,\mb{M_V}^2) = \int \frac{d^D p \, d^D q}{(2 \pi)^{2 D} \mu^{2 (D-4)}} \frac{1}{p^2+|\mb{M_\Phi}|^2} \frac{1}{(p+q)^2+|\mb{M_\Phi}|^2} \frac{1}{q^2+\mb{M_V}^2} .
\ee
This function is commonplace in two-loop vacuum calculations and corresponds to the scalar loop shown in \Fig{fig:loop}.  In $D = 4$ dimensions, this integral involves various divergences and subdivergences which must be appropriately regulated.  For the purposes of extracting sfermion soft masses, though, the physics of the regulator is irrelevant, since SUSY in the ultraviolet (UV) ensures finite sfermion soft masses.  We can therefore take the integral calculated using, say, minimal subtraction \cite{Ford:1991hw} and then expand in superspace.

Discarding terms that do not contribute to the final scalar masses, \Eq{eq:int} contains\footnote{Such discarded terms include single-logarithmic terms (which are scheme-dependent at two loops) and finite terms, which may be absorbed by a redefinition of couplings.}
\be
K_{2L} \supset \frac{q_\Phi^2 g'^2}{(4 \pi)^4} |\mb{M_\Phi}|^2 \left( 2 \mb{\Delta}  \log (\mb{\Delta}) \log \left({\tfrac{|\mb{M_\Phi}|^2}{\mu^2}}\right)  + (\mb{\Delta}+2) \log^2 \left({\tfrac{|\mb{M_\Phi}|^2}{\mu^2}}\right) + \Omega (\mb{\Delta})\right], \qquad \mb{\Delta} \equiv \frac{\mb{M_V}^2}{|\mb{M_\Phi}|^2}.
\label{eq:int2}
\ee
The dependence on the renormalization scale $\mu$ will drop out when we extract the soft masses.  Integral expressions for the function $\Omega(\mb{\Delta})$ appear in \Ref{Ford:1991hw}.   We can express $\Omega(\mb{\Delta})$ directly using dilogarithms as
\be
\Omega (\mb{\Delta}) = \sqrt{\mb{\Delta} (\mb{\Delta}-4)} \left(2 \zeta (2) + \log^2 \left(\mb{\alpha} \right) + 4 \text{Li}_2 \left[-\mb{\alpha} \right] \right) \mbox{~~~~with $\mb{\alpha} = \left(\sqrt{\frac{\mb{\Delta}}{4}} +\sqrt{\frac{\mb{\Delta}}{4} -1} \right)^{-2}$}.
\label{eq:omega}
\ee
  To find the final expression for the sfermion soft masses, we simply need to expand \Eq{eq:int2} to first order in $|\mb{q}|^2$ and integrate over superspace.  The resulting two-loop sfermion masses are
\be
\label{eq:AbelianFinal}
\widetilde{m}_q^2 = q_q^2 q^2_\Phi   \left(\frac{\alpha'}{2 \pi} \right )^2 \left | \frac{F}{M} \right |^2  f(\delta), \qquad \delta \equiv \frac{M_V^2}{M^2},
\ee
where
\be
f(\delta) = 2 \frac{\delta (4-\delta) ((4-\delta) +(\delta+2) \log(\delta))+2 (\delta-1) \Omega(\delta)}{\delta (4-\delta)^3},
\label{eq:fd}
\ee
and $\Omega(\delta)$ is defined in \Eq{eq:omega}.  Despite appearances, the function $f(\delta)$ is finite and real-valued for all positive values of $\delta$, including the region near $\delta = 4$.  \Eq{eq:AbelianFinal} is a key result of this work.

\subsection{Consistency Checks}
\label{sec:consistency}

The two-loop soft mass in \Eq{eq:AbelianFinal} satisfies a number of consistency checks.  First, this formula exhibits decoupling.  As the vector multiplet mass $M_V$ increases, we expect the sfermion soft masses to approach zero, such that $f(\delta) \to 0$ as $\delta \to \infty$.  The asymptotic behavior of $f(\delta)$ is
\be
\lim_{\delta \to \infty}  f(\delta)= 2 \frac{\log(\delta)-1}{\delta}.
\label{eq:larged}
\ee
This agrees with the asymptotic behavior found in \Ref{Gorbatov:2008qa} and satisfies our expectation that the soft masses vanish if the gauge superfield is completely decoupled.  

Second, we expect to recover the usual gauge-mediated results if the the gauge symmetry is restored, such that 
$f(\delta) \to 1$ as $\delta \to 0$.  The limiting behavior of $f(\delta)$ is
\be
\lim_{\delta \to 0}  f(\delta)= 1+\frac{\delta}{3} \left( \log (\delta)-\frac{1}{6} \right),
\label{eq:smalld}
\ee
also in agreement with \Ref{Gorbatov:2008qa}.  From this equation, one can see that the unbroken gauge mediation result is obtained in the unbroken limit, and that a Higgsing scale much below the messenger mass scale results in only a small suppression of the gauge mediated soft masses.  

\begin{figure}[t]
\centering
\includegraphics[height=3.0in]{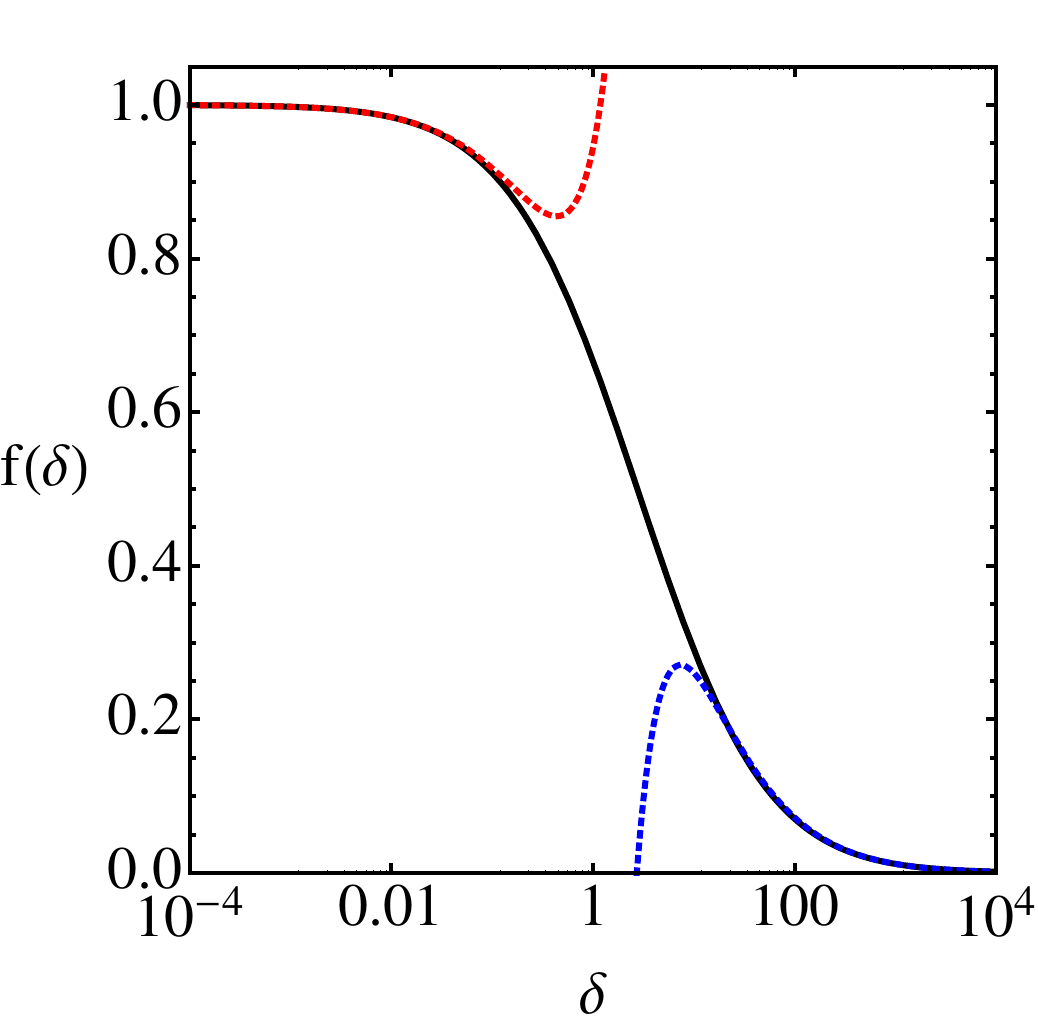}
\caption{The suppression of the scalar soft masses due to breaking of the mediating gauge group, as a function of $\delta \equiv M_V^2 / M^2$.  The full solution from \Eq{eq:fd} is plotted in black and the asymptotic forms are plotted in dotted blue for \Eq{eq:larged} and dotted red for \Eq{eq:smalld}.  One can see that if the gauge group is broken far below the messenger masses then the suppression of the soft masses is small, however if the gauge group is broken well above the messenger mass scale then the mediated soft masses are greatly suppressed.  Further, one can see that the approximate forms provide a good fit to the full solution in their respective limits, but break down rapidly in the crossover regime where $\delta \simeq 1$.}
\label{fig:f}
\end{figure}

In \Fig{fig:f}, we plot $f(\delta)$ alongside the two limiting expressions above.  One can see that while these expressions are valid in the respective limits, there is a large range $0.4 \lesssim \delta \lesssim 10$ in the crossover regime ($M_V \sim M$) for which these expressions do not provide a good approximation.

As a final consistency check, we can reverse the order of the superspace and momentum space integrals.  Expanding the integrand in \Eq{eq:int} to first order in $|\mb{q}|^2$ and then integrating over superspace, one expects to find a sum of terms which should correspond to the Feynman diagram integrand for the two-loop component field calculation, up to $\mathcal{O} (F^2/M^2)$.  Summing the ten diagrams calculated in \Ref{Gorbatov:2008qa}, we indeed find that the resulting integrands agree with the expanded version of \Eq{eq:int}, giving us confidence in the effective \kahler potential technique.  As expected, the resulting momentum-space integral is finite, demonstrating that although the integral \Eq{eq:int} is not finite, the resulting sfermion masses must be finite and independent of the renormalization scheme.  The expression in \Eq{eq:fd} is relatively simple compared to the previously found expressions in \Ref{Gorbatov:2008qa}, but the two final results agree numerically in the small $F/M^2$ limit.

\subsection{Non-Abelian Gauge Groups}
\label{sec:nonab}

In the Abelian case, we saw that a spurion analysis using the two-loop \kahler potential greatly simplified the calculation of soft masses in Higgsed gauge mediation.  This method also generalizes to the non-Abelian case.  Returning to the two-loop effective \kahler potential for non-Abelian gauge groups given in \Ref{Nibbelink:2005wc} and extending the model described in \Sec{sec:Kahler} to the non-Abelian case, the relevant terms in the \kahler potential are 
\be
\label{eq:preK2nonAbelian}
K_{2L} = -2 g'^2 \sum_{ab} \Tr [ t_\Phi^a t_\Phi^b]  \, I(\mb{M_\Phi}^2,\mb{M_\Phi}^2,\mb{M_V}^2)^{ba},
\ee
where $a,b$ label the generators of the the group, $t_\Phi$ are generators in the representation of the messenger field $\mb{\Phi}$, $\mb{M_V}^2$ is a matrix of gauge boson masses with entries
\be
\left(\mb{M_V}^2 \right)^{ab} = \left({M_V}^2 \right)^{ab} + g'^2 \mb{q}^\dagger (t_q^a t_q^b+t_q^b t_q^a) \mb{q},
\ee
and $t_q$ are generators in the representation of the visible field $\mb{q}$.  The $a$,$b$ indices on $I$ arise because $I$ is now a matrix-valued function of the gauge boson mass matrix.

We can simplify the expression in \Eq{eq:preK2nonAbelian} by employing the relation $\Tr [ t_\Phi^a t_\Phi^b] = C(\Phi) \delta^{ab}$ where $C(\Phi)$ is the Dynkin index of the messenger field representation.\footnote{In \Ref{Giudice:1998bp}, the messenger index for a given vector-like messenger pair is $n (\Phi) \equiv 2 C(\Phi)$.  This relation can be used to compare our results with those in the literature for unbroken gauge mediation.}  To evaluate the matrix-valued function $I$, it is simplest to work in the physical mass basis, where the gauge boson mass matrix has been diagonalized, i.e.\ ${D_V}^2 \equiv O^T {M_V}^2 O$ and the corresponding group generators have been rotated under the appropriate orthogonal transformation $T \equiv O^T t$.  Performing the same manipulations as in \Sec{sec:Kahler}, the visible sector scalar soft masses are
\be
\left(\widetilde{m}_q^2 \right)_{ij} = C(\Phi) \frac{\alpha'^2}{(2 \pi)^2}  \left | \frac{F}{M} \right |^2  \sum_a f(\delta^a) \, (T_q^a T_q^a)_{ij}, \qquad \delta^a \equiv \frac{{M^a_V}^2}{M^2},
\label{eq:nonab}
\ee
where $i,j$ are indices in the visible field representation, $f(\delta^a)$ is defined in \Eq{eq:fd}, and ${M^a_V}^2 = \left[{D_V}^2 \right]^{aa}$ is the physical gauge boson mass.  Using this simple formula, it is now possible to calculate the soft masses mediated by a non-Abelian gauge group with an arbitrary breaking pattern.  

If all gauge bosons have the same mass (i.e. $\delta^a = \delta$ for all $a$) we can simplify this expression by using $\sum_a [t_q^a t_q^a]_{ij} = C_2 (q) \delta_{ij}$, where $C_2(q)$ is the quadratic Casimir operator for the visible field representation.  In this limit,
\be
\widetilde{m}^2_q = C(\Phi) C_2 (q) \frac{\alpha'^2}{(2 \pi)^2}  \left | \frac{F}{M} \right |^2  f(\delta).
\ee
Note that for $\delta \to 0$ ($f(\delta) \to 1$), we recover the familiar results of unbroken gauge mediation.   

The extension to messengers charged under multiple groups is straightforward.  If we define $\widetilde{t}^a \equiv g^a t^a$ to include the gauge coupling constants, and define the rotation $\widetilde{T} \equiv O^T \widetilde{t}$ to diagonalize the full gauge boson mass matrix (including mixing between different gauge groups after Higgsing), then
\be
\label{eq:mostgeneral}
\left(\widetilde{m}^2_q \right)_{ij} = \frac{1}{64 \pi^4} \left | \frac{F}{M} \right |^2 \sum_a \Tr [\widetilde{T}_\Phi^a \widetilde{T}_\Phi^a]  \, f(\delta^a)  \,  (\widetilde{T}_q^a \widetilde{T}_q^a)_{ij},\qquad \delta^a \equiv \frac{{M^a_V}^2}{M^2},
\ee
where the $a$ sum runs over the physical gauge bosons.  We can generalize to multiple messenger fields simply by adding a sum over the states $\mb{\Phi}$.  The expression in \Eq{eq:mostgeneral} is very general and opens doors to a number of new model-building avenues which have previously evaded attention.  We will explore some of these possibilities in the next section.

\section{Examples}
\label{sec:examples}

Whenever the gauge group is broken far below the messenger mass scale, the suppression of the generated sfermion masses is negligible.  Although it might be interesting to study corrections to standard gauge-mediated soft masses due to electroweak symmetry breaking (i.e. non-zero $M_W$ and $M_Z$), these effects should be very small and thus not of interest here.\footnote{Similarly, our results are relevant for calculating the two-loop soft masses in deconstructed gaugino mediation \cite{Auzzi:2010mb,Auzzi:2010xc,Auzzi:2011gh,Auzzi:2011wt,Auzzi:2011eu}, though various three-loop contributions involving SSM gauginos and bifundamental link fields tend to give the dominant contribution unless the gauge-breaking scale is within a loop factor of the messenger scale.}  However, there are a number of scenarios involving gauge interactions beyond the SM gauge groups that are of interest, and we will outline three cases below:  (a) threshold corrections in  $SU(5)$ GUTs; (b) soft masses from deconstructed GUTs; and (c) extended gauge symmetries acting on the Higgs.  We discuss our key example of natural SUSY from flavor mediation in \Sec{sec:exampleFlav}.

\subsection{GUT Thresholds}
\label{sec:exampleGUT}

In order to maintain SUSY gauge coupling unification, messenger superfields are often arranged in complete GUT multiplets.  Usually, however, one only considers the generation of soft masses due to the unbroken SM gauge groups, discarding terms arising from loops of the massive GUT gauge bosons.  Here we emphasize that if the messenger masses are close to the GUT scale, then these extra contributions could be sizable (although still subdominant).

As an illustrative example, we study soft masses in an $SU(5)$ GUT.  If the soft masses are generated close to the GUT scale, we can assume that all gauge couplings are unified into a single coupling.  We also assume the standard embedding of each SM matter generation into a $\mb{10}$ and $\mb{\overline{5}}$ of $SU(5)$.  Once we know the breaking pattern, it is then straightforward to employ \Eq{eq:nonab} using the generators for both representations.  

We break the GUT group with an adjoint Higgs in the direction of hypercharge \cite{Georgi:1974sy}, such that $SU(5)\rightarrow SU(3)_C\times SU(2)_L\times U(1)_Y$.  With this breaking pattern, all additional $X$ and $Y$ vector superfields attain a mass $M_V$.  If the messengers have mass $M$, then we can identify $\delta = M_V^2/M^2$ and it is straightforward to find the sfermion masses using \Eq{eq:nonab}:
\be
\widetilde{m}^2 = C(\Phi) \, {C_2}_{\eff} \, \frac{\alpha_G^2}{(2 \pi)^2} \left | \frac{F}{M} \right|^2,
\ee
where
\be
\label{eq:naiveGUT}
\begin{array}{ccccl}
\alpha_G^2 {C_2}_{\eff} (\widetilde{Q}) &  = & \alpha_G^2 \left( \frac{21}{10}+\frac{3}{2} f (\delta) \right) &  \Leftrightarrow & \frac{3^2-1}{6} \alpha_S^2  + \frac{2^2-1}{4} \alpha_W^2 + \left(\frac{1}{6}\right)^2  \frac{5}{3} \alpha_Y^2  +\frac{3}{2} \alpha_G^2 f (\delta),  \\ \\
\alpha_G^2 {C_2}_{\eff} (\widetilde{U}^c) & = & \alpha_G^2 (\frac{8}{5}+2 f (\delta)) &  \Leftrightarrow & \frac{3^2-1}{6} \alpha_S^2 + \left( \frac{2}{3}\right)^2  \frac{5}{3} \alpha_Y^2 +2 \alpha_G^2 f (\delta),  \\ \\
\alpha_G^2 {C_2}_{\eff} (\widetilde{D}^c) & = & \alpha_G^2 (\frac{7}{5}+ f (\delta)) &  \Leftrightarrow &  \frac{3^2-1}{6} \alpha_S^2 + \left( \frac{1}{3}\right)^2 \frac{5}{3}  \alpha_Y^2  + \alpha_G^2 f (\delta),  \\ \\
\alpha_G^2 {C_2}_{\eff} (\widetilde{L}) & = & \alpha_G^2 (\frac{9}{10}+\frac{3}{2} f (\delta))  & \Leftrightarrow &  \frac{2^2-1}{4} \alpha_W^2 + \left( \frac{1}{2}\right)^2 \frac{5}{3}  \alpha_Y^2  +\frac{3}{2} \alpha_G^2 f (\delta),  \\ \\
\alpha_G^2 {C_2}_{\eff} (\widetilde{E}^c) & =  & \alpha_G^2 (\frac{3}{5}+3 f (\delta))  & \Leftrightarrow &  \frac{5}{3} \alpha_Y^2  +3 \alpha_G^2 f (\delta).
\end{array}
\ee
On the left, we show the contributions using the pure $SU(5)$ GUT calculation, whereas on the right, we show how these match onto the usual gauge-mediated contributions from the SM gauge group plus an additional threshold correction.  As expected, the contributions from the unbroken generators correspond to the usual gauge-mediated pattern when gauge couplings are unified, i.e.~$\alpha_S=\alpha_W=\frac{5}{3} \alpha_Y=\alpha_G$.  Hence, for messenger masses well below the GUT scale, one can evolve these contributions using the appropriate gauge couplings.  The threshold contributions from the massive $X$ and $Y$ gauge superfields take a rather simple form, depending only on $f(\delta)$.

Of course, in conventional scenarios with gauge-mediated SUSY breaking, the suppression of Planck-scale flavor-violating effects leads us to favor messenger masses $M \lesssim 10^{15}$ GeV, such that $\delta \gtrsim 10^{2}$.  Thus,  the threshold contributions from GUT breaking represent at most a $5 - 10 \%$ correction.  However, it is easy to envision scenarios for which the correction could be far more significant, for example, if Planck-mediated effects are suppressed by sequestering, or if the GUT scale is lowered due to deflected unification. Other concerns associated with high-scale GMSB may readily be ameliorated; for example, hadronic decays of the NLSP during Big Bang Nucleosynthesis can be forbidden by using multiple SUSY breaking as in \Ref{Cheung:2010mc} or gravitino decoupling as in \Refs{Luty:2002ff, Craig:2008vs}. 

\subsection{Deconstructed Orbifold GUTs}

Grand unified contributions to gauge mediation may prove even more significant when the GUT group is broken by dynamics {\it other} than the expectation value of an adjoint scalar Higgs.   For example, GUTs can be broken by boundary conditions in extra dimensions, or by the mixing of gauge groups in four dimensions, both of which can lead to larger effects from Higgsed gauge mediation.

To illustrate these possibilities, consider the simplest four-dimensional analogue of orbifold GUT breaking: a two-site deconstruction with groups $G_A = SU(5)$ and $G_B = SU(3) \times SU(2) \times U(1)$ \cite{Csaki:2001qm}. We introduce bifundamental link fields $\Sigma \oplus \overline \Sigma$ transforming as a $({\bf 5},{\bf \bar 5}) \oplus ({\bf \bar 5}, {\bf 5})$ whose vevs $\langle \Sigma \rangle \simeq \langle \overline \Sigma \rangle \equiv v$ break $SU(5)_A \times [SU(3) \times SU(2) \times U(1)]_B \to SU(3)_C \times SU(2)_L \times U(1)_Y$. Above the scale of Higgsing, the theory includes: $SU(3) \times SU(2) \times U(1)$ gauge bosons of $G_A$, denoted $A_i$ (where $i = 1,2,3$ runs over the three gauge groups); the $SU(5) / (SU(3) \times SU(2) \times U(1))$ gauge bosons of $G_A$, denoted $X,Y$; and the $SU(3) \times SU(2) \times U(1)$ gauge bosons of $G_B$, denoted $B_i$.  

At the scale of Higgsing, SM gauge couplings are determined via
\begin{equation}
\frac{1}{\alpha_i (v)} = \frac{1}{\alpha_A(v)} + \frac{1}{\alpha_{B_i}(v)},
\end{equation}
and it is useful to define the following combinations of gauge couplings:
\begin{equation}
g_i = \frac{g_A g_{B_i}}{\sqrt{g_A^2 + g_{B_i}^2}}, \qquad s_i = \sin \theta_i  \equiv \frac{g_A}{\sqrt{g_A^2 + g_{B_i}^2}}, \qquad c_i = \cos \theta_i \equiv \frac{g_{B_i}}{\sqrt{g_A^2 + g_{B_i}^2}}.
\end{equation}
There are three sets of gauge bosons below the scale $v$: one set of massless $SU(3)_C \times SU(2)_L \times U(1)_Y$ SM gauge bosons $V_i$ defined as $V_i = c_i A_i + s_i B_i  $; one set of massive $SU(3) \times SU(2) \times U(1)$ gauge bosons $U_i$ defined by $ U_i = - s_i A_i + c_i B_i $; and the massive $X,Y$ gauge bosons, which are simply the original $X,Y$ gauge bosons of $G_A$. The masses of the $U_i$ gauge bosons are $M_{U_i}^2 = 2 (g_{A}^2 + g_{B_i}^2) v^2$, and the masses of the $X,Y$ gauge bosons are $M_{X,Y}^2 = 2 g_A^2 v^2$.

Consider now chiral superfields $\mb{\Psi}_{A}, \mb{\Psi}_B$ charged under $G_A, G_B$, respectively. The kinetic terms for $\mb{\Psi}_A$ and $\mb{\Psi}_B$ include couplings to the gauge bosons from interactions of the form
\be 
\int d^4 \theta \, \left( \mb{\Psi}_A^\dag e^{2 g_A (\sum_i A_i + X + Y)} \mb{\Psi}_A + \mb{\Psi}_B^\dag e^{2 g_{B_i} B_i} \mb{\Psi}_B \right), 
\ee
where we have written the matrix-valued gauge fields as, e.g., $A \equiv T_a A^a$, and the sum over $i$ is implicit in the second term. The couplings to the corresponding mass eigenstate gauge bosons are  
\be
 \int d^4 \theta \, \left( \mb{\Psi}_A^\dag e^{2 g_i (V_i -  U_i \tan\theta_i) +2 g_A ( X  + Y )} \mb{\Psi}_A + \mb{\Psi}_B^\dag e^{2 g_i (V_i +  U_i \cot \theta_i)} \mb{\Psi}_B \right).
\ee
Below the scale of Higgsing, both $\mb{\Psi}_A$ and $\mb{\Psi}_B$ have canonical SM couplings to the massless gauge fields.  However, they also posses couplings to the massive gauge fields whose strength varies depending on the values of $g_A$ and $g_{B_i}$. The size of couplings to massive gauge bosons can considerably exceed couplings to the massless SM gauge bosons, such that the contributions from Higgsed gauge mediation can dominate. The couplings of fields charged under $G_A$ to the massive $X,Y$ and $U_i$ gauge bosons are given by
\be
g^A_{X,Y} = \frac{g_i}{\cos \theta_i} \quad \mbox{(for any $i$)}, \qquad g^A_{U_i} = - g_i \tan \theta_i,
\ee
while the couplings of those fields charged under $G_B$ are
\be
g^B_{X,Y} = 0,  \qquad g^B_{U_i} =  g_i \cot \theta_i.
\ee

The precise consequences for the spectrum of Higgsed gauge mediation depend on which gauge group the SSM and messenger superfields are coupled to in the UV.  Both the SSM and messenger superfields may be charged under either $G_A$ or $G_B$, or divided between the two; different configurations lead to different possible approaches to the problems of $SU(5)$ grand unification.  For example, if all SSM matter superfields are charged under $G_B$, there is no dimension-six proton decay from the exchange of $X,Y$ gauge bosons. Similarly, if the Higgs doublets are charged under $G_B$, the theory need not include color-triplet Higgs superfields, eliminating prohibitive dimension-five proton decay.  Alternatively, one could imagine a scenario in which the third-generation matter superfields are coupled to $G_A$, while the first two generations are coupled to $G_B$; this yields a successful prediction for $b/\tau$ Yukawa unification without making any unsuccessful unification predictions for lighter fermions. Various possible combinations are well-motivated from the perspective of unification physics.

The largest possible effects arise when the messengers and SSM superfields are charged under the same group.  Consider, for example, the case where all messenger and matter superfields are charged under $G_A$, and take for simplicity all $g_{B_i} = g_{B}$ ($\theta_i = \theta$) at the unification scale.  The soft masses are then given by
\be
\widetilde{m}^2 = C(\Phi) \, {C_2}_{\eff} \, \frac{\alpha_G^2}{(2 \pi)^2} \left | \frac{F}{M} \right|^2,
\ee
where now
\be
\begin{array}{rll}
{C_2}_{\eff} (\widetilde{Q})  & = &  \frac{21}{10}+ \frac{21}{10} t^4_\theta f(\delta / s^2_\theta) + \frac{3}{2 c^4_\theta} f (\delta),   \\ \\
{C_2}_{\eff} (\widetilde{U}^c)  & = & \frac{8}{5}+ \frac{8}{5}  t^4_\theta f(\delta / s^2_\theta) + \frac{2}{c^4_\theta} f (\delta),  \\ \\
{C_2}_{\eff} (\widetilde{D}^c)  & = & \frac{7}{5}+ \frac{7}{5}  t^4_\theta f(\delta / s^2_\theta) + \frac{1}{c^4_\theta} f (\delta), \\ \\
{C_2}_{\eff} (\widetilde{L})  & = & \frac{9}{10}+ \frac{9}{10}  t^4_\theta f(\delta / s^2_\theta) + \frac{3}{2c^4_\theta} f (\delta),    \\ \\
{C_2}_{\eff} (\widetilde{E}^c)  & =  & \frac{3}{5}+ \frac{3}{5}  t^4_\theta f(\delta / s^2_\theta) + \frac{3}{c^4_\theta} f (\delta).
\end{array}
\ee
Here, $\delta = M_{X,Y}^2/M^2$ and $t_\theta \equiv s_\theta / c_\theta$.  The three contributions to each soft mass arise from the massless SM gauge fields, the massive $U_i$ gauge fields, and the massive $X,Y$ gauge fields, respectively.  Note that the $X,Y$ contributions are enhanced over the na\"{i}ve GUT expectation in \Eq{eq:naiveGUT} for all values of $\theta$, while the $U_i$ contributions are new and parametrically enhanced by $\tan^4 \theta$.  Hence, the contributions of massive gauge bosons to sfermion masses may dominate over those of the massless fields for appropriate choices of the UV gauge couplings. These contributions may also introduce generation-dependent soft masses if SM families are split between $G_A$ and $G_B$, much as in \Ref{Craig:2011yk}.

\subsection{Extended Gauge Symmetries in the Higgs Sector}
\label{sec:exampleHiggs}

The mediation of SUSY breaking by spontaneously-broken extended gauge symmetries frequently arises in theories that modify the gauge interactions of the SSM Higgs multiplets.  The little hierarchy problem has motivated extensions of the SUSY Higgs sector \cite{Dine:2007xi}, and with recent excesses in Higgs-sensitive channels at the ATLAS and CMS experiments that could plausibly point towards $m_h \simeq 125$ GeV \cite{ATLAS-CONF-2011-163, CMS-PAS-HIG-11-032}, naturalness considerations only strengthen the motivation for such extensions.  These extensions broadly fall into two classes that, loosely speaking, correspond to $F$-term and $D$-term enhancements of the Higgs quartic.  The first class may be realized by introducing new superpotential couplings between the Higgs doublets and extra singlets or weak triplets.  These interactions increase the Higgs quartic coupling and hence the Higgs mass.  The second class may be realized by extending the SM gauge group.  If these new gauge symmetries are broken not far above the weak scale, and the fields responsible for Higgsing have comparable soft masses to this breaking scale, then the gauge $D$-terms do not decouple and the Higgs quartic coupling can be raised significantly. This again allows the Higgs mass to be raised without requiring unnaturally heavy stops \cite{Batra:2003nj,Maloney:2004rc}, although care must be taken to limit additional tree-level and one-loop contributions to the Higgs soft masses.

The extra gauge symmetries required by such non-decoupling $D$-terms might then also play the role of mediating SUSY breaking to the SSM.  This scenario represents a well-motivated example of Higgsed gauge mediation, where in this case the gauge symmetry is Higgsed just above the weak scale and the Higgsed gauge mediation sector might become accessible at LHC energies.  The number of models realizing such a scenario is potentially very large and so we merely demonstrate this scenario with the simple $U(1)_X$ example from \Ref{Batra:2003nj}.  In this example, the visible sector superfields $\{\mb{Q,U^c,D^c,L,E^c,H_u,H_d}\}$ have charges $\{0,-\frac{1}{2},+\frac{1}{2},0,+\frac{1}{2},+\frac{1}{2},-\frac{1}{2}\},$ respectively.  We envisage additional degrees of freedom that Higgs this extended gauge symmetry at the TeV scale, generating a mass for the vector superfield $M_V$.\footnote{If the spontaneous breaking of $U(1)_X$ is driven primarily by nonsupersymmetric terms in the scalar potential, then vevs deviating from $D$-flat directions can lead to tree-level corrections to the Higgs potential of order the Higgsing scale ($\sim$ TeV), potentially spoiling naturalness \cite{Batra:2003nj}. Thus we must take care to require that the Higgsing occurs in an approximately $D$-flat direction, which may be guaranteed if spontaneous symmetry breaking is driven by supersymmetric terms in the potential. Note also that these corrections only arise at one loop, rather than tree level, in non-Abelian gauge extensions.}  Now, if we add messenger fields $\mb{\Phi}/\mb{\Phi^c}$ with charges $\pm 1$ under $U(1)_X$ and couple them to a SUSY-breaking spurion $\mb{X}$ in the superpotential, then all sfermions charged under $U(1)_X$ will obtain soft masses at two loops.\footnote{These messengers do not break the gauge symmetry, and we assume a preserved messenger parity.  This is to avoid large tree-level corrections to the Higgs potential of order the SUSY-breaking scale.}  In this model, the generated soft masses are
\begin{eqnarray}
\widetilde{m}_{Q,L}^2 & = & 0, \\
\widetilde{m}_{U^c,D^c,E^c,H_u,H_d}^2 & = & \frac{\alpha_X^2}{(4 \pi)^2}  \left | \frac{F}{M} \right |^2  f(\delta), \qquad \delta \equiv \frac{M_V^2}{M^2}.
\end{eqnarray}
In this case, one would expect $M_V \ll M$ and hence $f(\delta) \approx 1+\frac{\delta}{3} \left( \log (\delta)-\frac{1}{6} \right)$ as in \Eq{eq:smalld}. Of course, we must also include conventional messengers charged under the SM gauge group in order to generate masses for SSM gauginos and the left-handed squarks and sleptons, but including these ingredients does not spoil the Higgsed spectrum.  If extended gauge symmetries play any role in raising the Higgs mass, they may leave their signature in modifications to the SUSY soft-mass spectrum arising from Higgsed gauge mediation.

We should note an important subtlety in the above calculation.  In order for the non-decoupling $D$-terms to be relevant to the little hierarchy problem, it is necessary that the gauge symmetry breaking also breaks SUSY, i.e.~$\mb{M_V^2} = M_V^2 (1+\theta^4 \overline{m}^2)$ where $M_V\sim \overline{m} \sim$ TeV.  As the gauge symmetry breaking threshold is not supersymmetric, it might seem that this spoils the calculation of the soft masses.  Indeed, there are corrections to the result stated, but we may show that they are subdominant whenever the gauge interactions are in the perturbative regime.  We now have two SUSY-breaking spurions:  $\mb{X^\dagger X}=M^2 |1+\theta^2 F/M|^2$ and $\mb{M_V^2}=M_V^2 (1+\theta^4 \overline{m}^2)$.  Expanding whichever terms might arise in the \kahler potential, we find soft masses proportional to either $|F/M|^2$ or $\overline{m}^2$.  If the gauge interactions are perturbative, the coefficient of the $|F/M|^2$ term typically arises at two loops $\simeq \alpha_X^2/(2 \pi)^2$, and the coefficient of the $\overline{m}^2$ term can arise at one loop $\simeq \alpha_X/2 \pi$.  Now, for gauge mediation to be applicable we require 
\be
\left( \frac{\alpha_X}{2 \pi}\right)^2 \left|\frac{F}{M} \right|^2 \sim \text{TeV},
\ee
and for non-decoupling $D$-terms to be relevant we require $\overline{m}\sim$ TeV.  Since for even $\mathcal{O}(1)$ gauge couplings we have $\alpha_X/2 \pi \sim 10^{-2}$, we therefore expect corrections due to the non-supersymmetric nature of the gauge threshold to be subdominant.

\section{Flavor Mediation}
\label{sec:exampleFlav}

Our key example where Higgsed gauge mediation may play a crucial role is ``flavor mediation'', where global flavor symmetries are gauged and soft masses are mediated to squarks and sleptons via these gauged flavor symmetries.   Consideration of the full breadth of this scenario is beyond the scope and objectives of this paper.  Here, we outline some key features in order to demonstrate this particularly appealing application of Higgsed gauge mediation, returning to the construction of a complete model in future work.

The general structure of a flavor mediation model is as follows.  We gauge some flavor subgroup of the SSM matter, say $SU(3)_F$.\footnote{This $SU(3)_F$ could be the left-handed quark symmetry $SU(3)_Q$, right-handed up-type quark symmetry $SU(3)_U$, right-handed down-type quark symmetry $SU(3)_D$, or various diagonal combinations.  It could alternatively act on leptons.}  If necessary, additional matter content must be added to cancel SM gauge anomalies depending on which flavor symmetry is gauged.   In addition, we envision vector-like messengers $\mb{\Phi}/\mb{\Phi^c}$ in some representation of $SU(3)_F$ with a superpotential coupling to the SUSY-breaking spurion $\mb{X}$.  Through these messengers, any fields transforming under $SU(3)_F$ will feel SUSY breaking via gauge mediation.\footnote{Once again, SSM gaugino masses can be generated in the standard manner if there are also messenger fields charged under the SM gauge group. Of course, this generates conventional gauge-mediated soft masses for all three generations, but this need not spoil the hierarchy introduced by flavor mediation.}  In particular, scalars contained in chiral supermultiplets charged under the gauged flavor symmetry will obtain soft masses at two loops.  This is not the full story, however, as the flavor symmetry must be spontaneously broken in order to generate the SSM Yukawa couplings.  Moreover, the flavor gauge bosons typically obtain masses correlated with the pattern of Yukawa couplings.  As we will see, the spontaneous breaking of the flavor symmetry feeds into SSM sfermion mass structures via Higgsed gauge mediation in an intriguing way.

As a toy example, we consider symmetry breaking by three $SU(3)_F$ fundamentals $\mb{\Psi}_{ij}$, with a hierarchical pattern of vevs, in analogy with SSM Yukawas.  We may diagonalize this $3 \times 3$ matrix via flavor rotations into the form
\be
\langle \Psi \rangle = \left( \begin{array}{ccc}
v_3 & 0 & 0 \\
0 & v_2 & 0 \\
0 & 0 & v_1 \end{array} \right),
\ee
where $v_3\gg v_2\gg v_1$.  This vev completely breaks $SU(3)_F$, however it is more instructive to picture the symmetry breaking pattern schematically as
\be
v_3 : SU(3)_F \rightarrow SU(2)_F
\ee
followed by
\be
v_2 : SU(2)_F \rightarrow \text{nothing}.
\ee
The crucial feature is that $v_1$ is not necessary to break any residual $U(1)_F$ symmetries, and this fact turns out to be highly appealing for the sfermion soft mass structure.

We now turn to the soft masses from Higgsed gauge mediation.   In the limit that $SU(3)_F$ is unbroken, then the soft masses for all sfermions are clearly degenerate.  When we turn on $v_3$, the corresponding third generation soft masses become suppressed due to the breaking of the mediating gauge group, yielding $f(\delta) < 1$.  Thus, generations with \emph{large} Yukawas acquire \emph{suppressed} soft masses.  Now if we turn on $v_2$, the gauge symmetry becomes fully broken, and the first two generation sfermions acquire \emph{the same} soft mass from mediation via the broken $SU(2)_F$.  As there is no remaining $U(1)_F$ flavor symmetry around, the first generation soft masses are not enhanced compared to the second generation via any additional gauge mediation, and the first two generations instead remain degenerate, up to a small contribution from $v_1$.

In this way, the mechanism of flavor mediation can \emph{automatically} give the features desired of a ``natural SUSY'' spectrum:\footnote{For models and constraints on natural SUSY see \Refs{Dine:1993np,Pouliot:1993zm,Barbieri:1995rs,Dimopoulos:1995mi,Pomarol:1995xc,Barbieri:1995uv,Cohen:1996vb,Barbieri:1997tu,Gabella:2007cp,Sundrum:2009gv,Barbieri:2010pd,Barbieri:2010ar,Craig:2011yk,Gherghetta:2011wc,Kats:2011qh,Papucci:2011wy,Brust:2011tb,Delgado:2011kr,Desai:2011th,Akula:2011jx,Ajaib:2011hs,Ishiwata:2011fu,Lodone:2011pv,He:2011tp,Arvanitaki:2011ck,Auzzi:2011eu,Csaki:2012fh}.  For recent constructions involving gauged flavor symmetries see \Ref{Grinstein:2010ve}.}
\begin{itemize}
\item  Suppressed soft masses for the third-generation sfermions;
\item  Degenerate first- and second-generation sfermions, respecting an approximate $U(2)$ symmetry.
\end{itemize}
These two features follow directly from the combination of the flavor symmetry structure and the mechanism of Higgsed gauge mediation, and it is not necessary to impose an ad hoc $U(2)$ flavor symmetry or to treat the first two generations differently from the third as is often required in models of natural SUSY.\footnote{We note that one could also consider examples where $SU(2)$ subgroups are gauged in order to generate additional degenerate contributions to the soft masses of the first two generations, however an explanation for preferential treatment of the first two generations would be lacking.  Such problems are avoided here by gauging a full $SU(3)_F$ flavor symmetry, thus treating all generations equally.  Of course, our mechanism would not work as successfully with a $U(3)_F$ flavor symmetry, owing to the appearance of an extra $U(1)_F$ gauge boson that would get its mass from $v_1$.  This framework also includes the gauging of Froggatt-Nielsen $U(1)$ symmetries \cite{Kaplan:1998jk,Kaplan:1999iq} as a subclass.  In this case, the soft mass will be proportional to the charge-squared under this symmetry, whereas Yukawas will be largest whenever the charge is small, again obtaining a natural SUSY spectrum.}

\begin{figure}[t]
\centering
\includegraphics[height=3.05in]{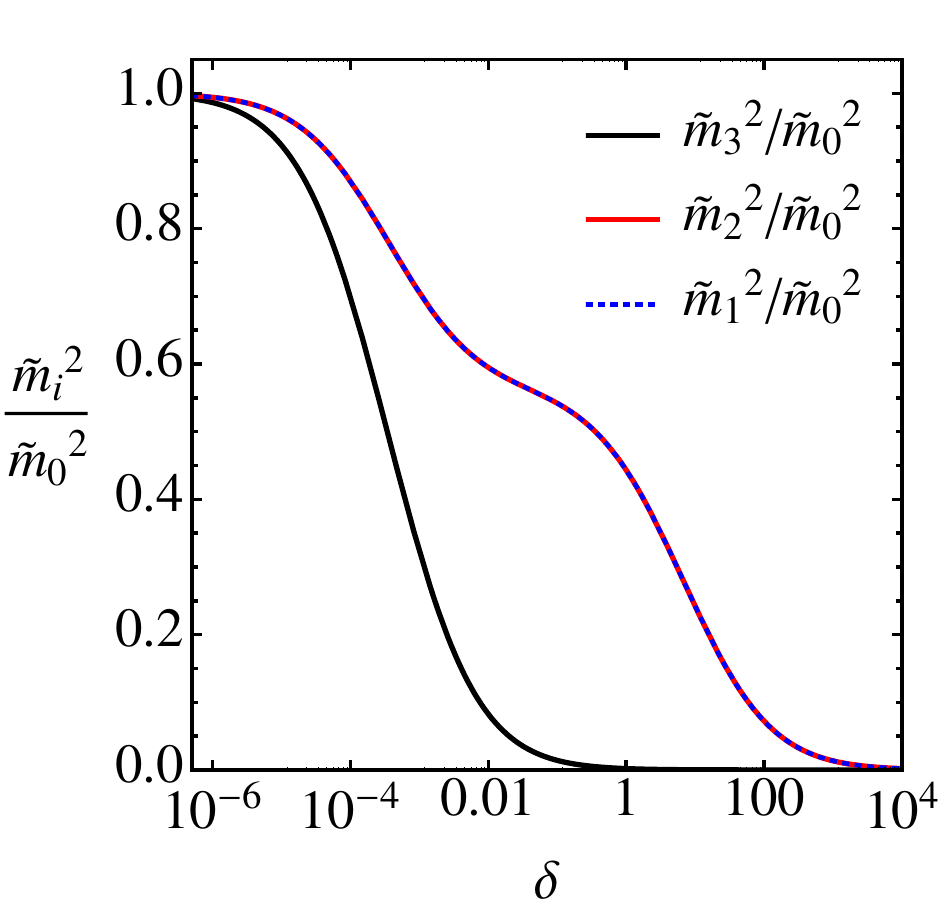}  \hspace{0.5in} \includegraphics[height=3.05in]{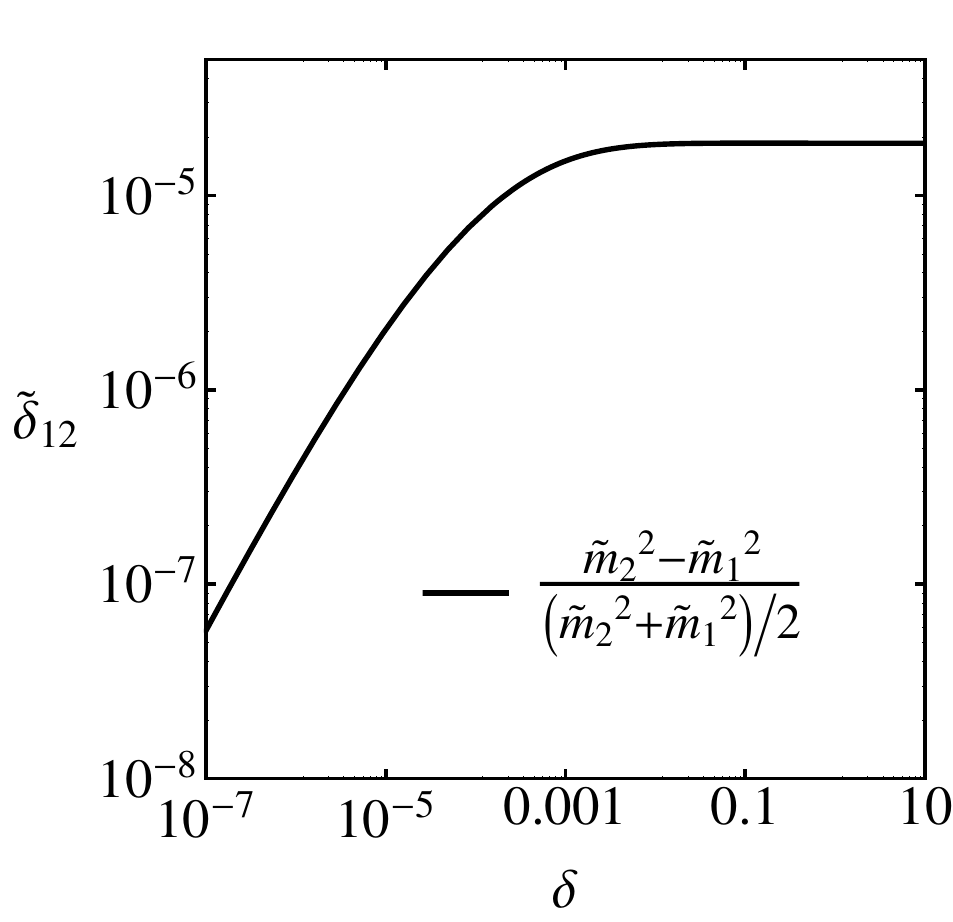}
\caption{Left panel: the spectrum of soft masses generated in a toy model of flavor mediation with vevs $v_3 = 173 \, v$, $v_2 = 1.29 \, v$ and $v_1 = 2.5 \times 10^{-3} \, v$.  The masses are plotted as a function of $\delta = g_F^2 v^2/M^2$ and are shown relative to the soft masses which would be generated in the unbroken limit, denoted $\widetilde{m}_0^2$.  The third generation is plotted in black, second in red, and first in blue dotted.  The suppression of the third-generation soft mass relative to the first two generations is clear, and the high degree of degeneracy between the first two generations is also demonstrated.  The extra contribution to the first two generations from the remaining $SU(2)$ can be seen, as well as their corresponding suppression whenever the breaking of this symmetry becomes important.  Right panel:  mass-squared splitting between the first two generations, where $\widetilde{\delta}_{12} = (\widetilde{m}_2^2-\widetilde{m}_1^2)/((\widetilde{m}_2^2+\widetilde{m}_1^2 )/2)$.}
\label{fig:flavmed}
\end{figure}

To give numerical support to these heuristic arguments, we calculate a particular spectrum in detail by employing the methods developed in \Sec{sec:HGM}.   Choosing the suggestive vevs $v_3 = 173 \, v$, $v_2 = 1.29 \, v$ and $v_1 = 2.5 \times 10^{-3} \, v$ (motivated, of course, by the up-type Yukawa structure), we compute the generated soft masses as a function of $\delta =g_F^2 v^2/M^2$ using \Eq{eq:nonab} and rotating to the gauge boson mass basis.  The results are shown in \Fig{fig:flavmed}.  For $\delta \approx 0.1$, the third-generation soft masses are suppressed in comparison to the other two generations.  We also show the mass-squared splittings, which make it clear that the first two generations are automatically highly degenerate, with mass-squared splittings of $\widetilde{\delta}_{12} \approx 2 \times 10^{-5}$, where $\widetilde{\delta}_{12} = (\widetilde{m}_2^2-\widetilde{m}_1^2)/((\widetilde{m}_2^2+\widetilde{m}_1^2 )/2)$, in the parameter region where the third-generation soft masses are suppressed.

The toy model outlined above does not constitute a complete model of flavor mediation, which would typically involve additional matter for anomaly cancellation, treatment of the full flavor symmetry group (including the breaking pattern, flavor-boson masses, and constraints on flavor-changing neutral currents), and consideration of the SUSY-breaking scale and cosmology.  Although oversimplified at present, it is apparent that this scenario provides an attractive framework in which to construct models of natural SUSY.

\section{Conclusion}
\label{sec:conclude}

Gauge mediation via SM gauge interactions has long provided an attractive framework for SUSY breaking in the SSM due to the calculability and universality of the resulting soft mass spectrum. However, this picture appears disfavored by the first few inverse femtobarns of LHC data. Increasingly stringent direct limits on the production of first-generation squarks have pushed their masses above $\sim$ 1 TeV, imperiling the naturalness of a flavor-universal soft spectrum. Retaining supersymmetric naturalness in light of data instead requires additional dynamics to distinguish the third generation while preserving an approximate symmetry between the first two. Moreover, hints of a possible Higgs boson near 125 GeV are particularly hard to accommodate in conventional gauge mediation \cite{Draper:2011hd}. 

This strongly motivates generalizations of gauge mediation involving additional spontaneously-broken gauge symmetries that predict deviations from the conventional soft spectrum. Although the effects on sparticle masses of Higgsing the mediating gauge group have been computed previously by direct evaluation of Feynman diagrams \cite{Gorbatov:2008qa}, their form remained somewhat baroque. In this paper, we have presented a much more compact expression for the soft spectrum, obtained by a spurion analysis of the two-loop effective \kahler potential. The concise form of soft masses in \Eq{eq:mostgeneral} is entirely general, up to corrections of $\mathcal{O}(F/M^2)$, and invites application to a wide variety of models.

To this end, we have explored several well-motivated gauge extensions of the SSM that may realize Higgsed gauge mediation. Even the simplest GUT scenarios possess additional massive gauge bosons that alter conventional gauge-mediated predictions if the messenger and unifications scales are not widely separated. These effects may be particularly strong in four-dimensional realizations of orbifold GUT symmetry breaking, in which SSM fields may couple preferentially to massive gauge bosons.  In models with extended gauge symmetries in the Higgs sector---motivated by the possibility of increasing the Higgs boson mass through non-decoupling $D$-terms---Higgsed gauge mediation can substantially modify the sfermion spectrum.

Perhaps the most attractive possibility, from the perspective of naturalness and flavor, is the gauging of SM flavor symmetries. The pattern of spontaneous breaking of these symmetries necessarily mirrors the texture of SSM Yukawas, leading to a spectrum of soft masses that naturally distinguishes the third-generation scalars while effortlessly preserving a $U(2)$ symmetry between the first two---precisely the spectrum favored by naturalness, indirect flavor constraints, and direct search limits.  We have only briefly explored the features of such a model here, leaving a detailed study of its phenomenology and signatures to future work. 

Of course, these are but a few of many possible model-building applications of our results. Gauge extensions of the SSM are manifold, making it easy to envision that additional vector fields may exist above the weak scale and play a role in the communication of SUSY breaking. The interplay of mass scales apparent in the spectrum of Higgsed gauge mediation suggests a wide range of potential consequences for the masses of SSM scalars, as well as the rich phenomenology associated with new gauge degrees of freedom. If SUSY indeed plays a role in stabilizing the weak scale, such physics may soon be apparent at the LHC. 

\acknowledgements{We benefitted from conversations with Matthew Sudano.  NC is supported in part by the NSF under grant PHY-0907744 and gratefully acknowledges support from the Institute for Advanced Study.  MM and JT are supported by the U.S. Department of Energy (DOE) under cooperative research agreement DE-FG02-05ER-41360.  MM is also supported by a Simons Postdoctoral Fellowship, and JT is also supported by the DOE under the Early Career research program DE-FG02-11ER-41741.}

\bibliography{HGMref}

\end{document}